\documentclass[%
 reprint,
 nofootinbib,
 amsmath,amssymb,
 aps,
]{revtex4-2}

\usepackage{color}
\usepackage{graphicx}
\usepackage{dcolumn}
\usepackage{bm}
\usepackage{hyperref}

\newcommand{\gray}{$\gamma$-ray }
\newcommand{\sv}{\langle \sigma v \rangle}

\newcommand{\ajnvI}[1]{\textcolor{black}{#1}}

\newcommand{\DvI}[1]{\textcolor{black}{#1}}

\begin{document}

\preprint{APS/123-QED}

\title{
\ajnvI{
Constraining Dark matter annihilation with Dark Energy Survey Y3 LSBG sample}
}

\author{Daiki Hashimoto}
 \email{hashimoto.daiki@f.mbox.nagoya-u.ac.jp}
\affiliation{%
 Division of particle and astrophysical sciences, Graduate School of Science, Nagoya University, Furocho Chikusa, Nagoya, 464-8602, Aichi, Japan
}%

\author{Atsushi J. Nishizawa}%
\affiliation{
 Institute for Advanced Research, Nagoya University, Furocho Chikusa, Nagoya, 464-8602, Aichi, Japan \\
 Division of particle and astrophysical sciences, Graduate School of Science, Nagoya University, Furocho Chikusa, Nagoya, 464-8602, Aichi, Japan
}%

\author{Masahiro Takada}%
\affiliation{%
 Kavli Institute for the Physics and Mathematics of the Universe (WPI), The University of Tokyo Institutes for Advanced Study (UTIAS),\\~~~The University of Tokyo, 5-1-5 Kashiwanoha, Kashiwa-shi, Chiba, 277-8583, Japan
}%

\date{\today}

\begin{abstract}
To reveal natures of the dark matter~(DM) particles, 
a \gray signal produced in annihilation processes of DM into the standard model particles has been one of the major \ajnvI{probes}.
\ajnvI{The}
cross-correlation 
\ajnvI{between}
highly DM dominated structures, such as local dwarf galaxies, and observed photons in the direction of the structures, has been 
\ajnvI{explored}
and provided stringent constraints on the annihilation rate. 
In our previous work, we have 
\ajnvI{shown that it is sufficient to know the distance distribution of the galaxy sample and individual distance measurement is not required for constraining the annihilation rate.}
In this work, we apply 
\ajnvI{the}
method 
to low-surface brightness galaxies~(LSBGs) with unknown individual redshifts provided from the Dark Energy Survey~(DES) Year 3 data. With all DES-LSBGs of $\sim$24,000 objects, we find that the upper limits of the cross section $\sv$ for $b\bar{b}$ channel with 95\% C.L. is $\sim~3\times 10^{-25}~\rm cm^3/s$ at DM mass of 100~GeV. 
\ajnvI{To be more conservative, }
we remove $\sim$ 7000 LSBGs within $1^{\circ}$ from 
resolved \gray point sources and then the 
\ajnvI{constraint becomes}
$\sim$30 \% weaker than the one with all samples in all DM mass ranges.
\end{abstract}

\maketitle

\section{Introduction}
\label{sec:intro}

As one of the most promising dark matter~(DM) candidates, weakly interacting massive particles~(WIMPs)~\citep{1996PhR...267..195J}, which are considered to be produced in the early Universe in the thermal equilibrium with the standard-model~(SM) particles, have been searched in several decades.
Great efforts have been kept to search for signals originated from interactions of WIMPs with SM particles~\citep{2018EPJC...78..203A, 2019arXiv190407915L}. 
In the context of the astrophysics, researchers have focused on indirect searches for DM annihilation signals, particularly, emission of $\gamma$ rays, positrons and neutrinos produced by the annihilation of DM particles and secondary cascade processes~\citep[e.g.,][]{2005PhR...405..279B, 2018RPPh...81f6201R}.
In terms of the \gray signal, the cross correlation of observed \gray photons with any astronomical structure, such as local galaxies, galaxy groups, clusters, the Milky Way dwarf spheroidals~(MW dSphs) and the Galactic center, has been studied in numerous works~\citep[e.g.,][]{2014PhRvD..89f3515M, 2014PhRvD..90b3514A, 2015ApJ...812..159A, 2016PDU....12....1D, 2017ApJ...840...43A, 2020PhRvD.102d3012A, 2021A&A...648A..60A, 2021MNRAS.502.4039T}, and then upper limits on the annihilation cross-section has been provided.
The most robust and stringent constraint is provided by the cross-correlation analysis using 27 MW dSphs with unresolved \gray background~(UGRB), which is a residual photon-flux field produced by subtraction of the Galactic and isotropic emissions as well as resolved \gray point-source emissions from the photon data of the \textit{Fermi} Large Area Telescope~(LAT) observation, and the upper limit with 95\% C.L. is found at $\sv \sim 2\times 10^{-26}~\rm [cm^3/s]$ at DM mass of 100 GeV~\citep{2020JCAP...02..012H}.

In our previous work~\citep{2020JCAP...01..059H}, low-surface brightness galaxies~(LSBGs) have been proposed as target objects to probe the \gray signal (LSBGs have been used for the prove in a few works, such as Ref.~\citep{2021MNRAS.501.4238B}).
LSBG has several favorable aspects for the purpose such as being relative massive~\citep{2019MNRAS.484.4865P}, highly DM-dominated~\citep{2017MNRAS.470.1512W} and expected to be quiescent in $\gamma$ rays due to relatively quiescent star-forming activities~\citep{2019MNRAS.483.1754D}.
We have explored the annihilation \gray signal in the UGRB sky performing the joint likelihood analysis with known-redshift LSBGs, which is included in a tiny fraction of a LSBG catalog produced from the Hyper Suprime-Cam~(HSC) data~\citep{2018ApJ...857..104G}.
In our another previous work~\citep{2021arXiv210908832H}, we have presented the method for the signal probe using all LSBGs ($\sim 800$ objects) of the HSC-LSBG catalog.
In the method, we randomly assign object redshifts from the redshift distribution of the overall sample $dN/dz$, which is obtained by the so-called clustering redshift method, in which we measure angular cross-correlations of LSBGs and a spectroscopic redshift (spec-$z$) sample as a reference sample in different redshift bins, and convert the correlations to the $dN/dz$ amplitudes.
With the joint likelihood analysis using the full sample, we have shown that it is sufficient for the DM cross-section constraint to use $dN/dz$ of the overall catalog without measuring their individual redshifts.
Moreover, we have found that the upper limit on the cross section scales with an inverse of the number of objects $N$, rather than $\sqrt{N}$.

In this work, we apply our method to the Dark Energy Survey~(DES) LSBGs~\citep{2021ApJS..252...18T} including $\sim$24,000 objects in the sky coverage of $\sim$5000 $\rm deg^2$.
As a difference between procedures in this work and the previous work using the full HSC-LSBG sample, we consider the cross-covariance matrix of the angular cross-correlation between different redshift bins.
Moreover, because variability of flux-model parameters of bright sources nearby LSBGs can significantly change the putative flux of LSBGs, we consider a better UGRB-field production for the careful estimation of the putative flux.

This paper is organized as follows.
We describe datasets in our analysis in Section~\ref{sec:data}, and the method for production of the UGRB field and estimate of a likelihood profile of the putative flux at LSBG position in Section~\ref{sec:bg_flux}.
In Section~\ref{sec:met}, we revisit the methodology of our analysis for the $dN/dz$ measurement and composite likelihood and find the results in the composite analysis using the full sample of DES LSBGs in Section~\ref{sec:result}.
Finally, we summarize our study in Section~\ref{sec:summary}.

\section{Dataset}
\label{sec:data}
\subsection{LSBG sample}
\label{ssec:lsbg}
DES is an imaging survey of five 
\ajnvI{optical}
broad bands ($g,r,i,z$ and $Y$) 
covering $\sim 5000~\rm deg^2$ of the southern hemisphere using the Dark Energy Camera~(DECam)~\citep{2015AJ....150..150F} on the 4-m Blanco Telescope at the Cerro Tololo Inter-American Observatory.
The DECam composed by 74 CCDs with a central pixel scale of $0.263^{\prime \prime}$ has a wide field of view of $3~\rm deg^2$.
The single epoch processing pipeline \texttt{Final Cut}~\citep{2018PASP..130g4501M} provides reduced images and also performs source detection and measurement of galaxies with \texttt{SourceExtractor}~\citep{1996A&AS..117..393B}, which can discover faint objects.
In the Wide Survey of DES Year 3, the median coadd magnitude limits in the $g,r,i,z$ and $Y$ band with signal-to-noise ratio~(S/N) of 10 are 24.3, 24.0, 23.3, 22.6 and 21.4, respectively~\citep{2021ApJS..254...24S}.
For detection of 
\ajnvI{LSBGs},
\ajnvI{dedicated}
background estimation is indispensable ~\citep{2018PASP..130g4501M, 2021ApJS..254...24S}. 

\ajnvI{The }
LSBG 
\ajnvI{sample}
\footnote{\url{http://desdr-server.ncsa.illinois.edu/despublic/other_files/y3-lsbg/}}~\citep{2021ApJS..252...18T} 
\ajnvI{is constructed from the DES Y3 Gold sample as follows.}

~

\noindent 1) \textit{Extraction of candidates from Y3 Gold catalog}

\noindent From the DES Y3 Gold coadd 
catalog (v2.2)~\citep{2021ApJS..254...24S}, 
point-like objects \ajnvI{are removed} based on the \ajnvI{size, surface brightness, ellipticity and color cuts as}
\begin{align}
    -0.1 < g-i < 1.4 \notag \\
    g-r > 0.7\times (g-i) - 0.4 \notag \\
    g-r < 0.7\times (g-i) + 0.4. \notag
\end{align}
The initial catalog after source selections above contains 419,895 candidates \ajnvI{after this source selection}.

~

\noindent 2) \textit{Sample selection by machine learning algorithm}

\noindent A machine learning method is used to \ajnvI{further} remove \ajnvI{the} contamination 
such as giant elliptical galaxies, compact objects with the diffuse foreground or background 
and knots of large spiral galaxies.
\ajnvI{The training set is constructed by visual inspection of the image and 7760 objects are labeled. Out of that, 640 objects are labeled as LSBGs.}
\ajnvI{The machine learning classification identified 44,979 objects as the LSBG candidate. Note that the $\sim9\%$ LSBGs are not detected with this method.}

~

\noindent 3) \textit{Visual inspection \& S$\acute{e}$rsic model fitting}

\noindent The 
candidates have been visually inspected using the cutouts of $30^{\prime\prime} \times 30^{\prime\prime}$ centered at the candidate position of each of the candidates.
\ajnvI{Finally, the light profile is fitted to the S$\acute{e}$rsic profile to distinguish the LSBGs from other objects.}
\ajnvI{All the LSBGs should pass the selections of}
effective radii $R_{\rm eff}(g) > 2.5^{\prime\prime}$ and $\bar{\mu}_{\rm eff}(g) > 24.2~{\rm mag/arcsec^2}$.

~

The final catalog contains 23,790 LSBGs with the surface brightness fainter than 24.2 $\rm mag/arcsec^2$.
\ajnvI{They}
are divided into 7,805 red ($g-i \geq 0.6$) and 15,985 blue ($g-i < 0.6$) LSBGs.
While the red LSBGs are 
\ajnvI{strongly}
clustering, the blue \ajnvI{LSBGs are rather uniformly distributed.}
According to the stellar population model of \citep{2017ApJ...835...77M}, typical ages of red and blue LSBGs are 4~Gyr and 1~Gyr with [Fe/H] = -0.4, respectively.

\subsection{Spec-$z$ sample}
\label{ssec:specz}
To measure the $dN/dz$ of the DES LSBG catalog based on the \ajnvI{clustering} redshift 
method, we 
use the final data release (DR3) of 6-degree Field Galaxy Survey~(6dFGS) \ajnvI{spectroscopic redshift} sample \footnote{\url{http://www-wfau.roe.ac.uk/6dFGS/download.html}}~\citep{2004MNRAS.355..747J, 2009MNRAS.399..683J}, \ajnvI{which fully overlapped with the DES sky coverage and expected redshift range.}
\ajnvI{The limiting magnitudes} in $K, H$ and $J$ band are 12.65, 12.95 and 13.75, respectively, and \ajnvI{the catalog} consists of 1,447 
fields ($5.7^{\circ}$ field of view) of the telescope. 
In each field, 150 
spectra are obtained simultaneously by individual fibers which have 100 $\mu$m ($\sim 6.7^{\prime\prime}$)-fiber diameter size of each. Due to the fiber collision, spectra of neighboring objects closer than $5^{\prime}.7$ cannot be measured simultaneously.

In our analysis, we consider the 6dFGS sample in the DES Y1 Gold footprint~\citep{2018ApJS..235...33D} for the $dN/dz$ measurement and the redshift range of the sample is limited to $z\leq 0.15$.
This range would include redshifts of our LSBGs because H$\alpha$-selected LSBG samples reside within a few hundreds Mpc \citep{2019MNRAS.483.1754D, 2019ApJS..242...11L}.
As described in the DR3 report~\citep{2009MNRAS.399..683J}, we only use samples with the redshift-quality flag $Q=3,4$.

\subsection{LAT photon data}
\label{ssec:lat}
To probe \gray signal from the DM annihilation, we explore the \gray data by \textit{Fermi}-LAT 
\ajnvI{collected from Oct. 27, 2008 to July 6, 2021}
and 
\ajnvI{take}
the photon event class \texttt{P8R3 SOURCE} \DvI{and the instrument response function for the event class \texttt{P8R3 SOURCE V3}} as the photon count data for analyzing point sources recommended by the Fermi collaboration~\citep{2013arXiv1303.3514A}.
We select the photons satisfying the criteria of 
\texttt{DATAQUAL>0 \& LATCONFIG==1}.
In our work, we select sky regions including the survey footprint of the DES Y3, which are composed by 43 regions of interest~(ROIs). Every ROI is a patch of $20^{\circ}\times 20^{\circ}$ \ajnvI{square region} with $0.1^{\circ}$ spatial grid size 
\ajnvI{allowing 2.5 deg overlaps with the adjoining ROIs.} 
\ajnvI{In order to define the photon energy range to be used in our analysis, we need to consider the two distinct effects:}
in lower energy ranges, photons around bright sources likely leak \ajnvI{to the neighboring pixels} due to broadening of the LAT point-spread function~(PSF) while in higher energy ranges, the photon statistics decreases.
\ajnvI{Therefore, we limit the range of photon from 500 MeV to 500 GeV and within that range, the photon energy is further  binned in logarithmically equally spanned in 24.}
To avoid a contamination of photons produced by interactions of cosmic rays with the Earth's atmosphere, we exclude the photon data of zenith angles larger than $100^{\circ}$.
In the LAT data analysis, we use the open-source package \texttt{fermipy}(v1.0.1)~\citep{2017ICRC...35..824W},
which \ajnvI{is} based on the \texttt{Fermi Science Tools}(v2.0.8)~\citep{2019ascl.soft05011F}.

\section{UGRB Flux at the LSBG position}
\label{sec:bg_flux}

\subsection{UGRB construction}
\label{ssec:ugrb}

Before analyses for probing the DM annihilation signal, we construct the UGRB sky from the observed photon data, which needs to estimate contributions from bright \gray sources to the total emission in our ROIs. The bright sources are three types of sources; the diffuse emission of Galactic and isotropic component as well as resolved point sources by the LAT.
For modeling fluxes of these sources, we adopt the standard Galactic \texttt{(gll\_iem\_v07.fits)} and isotropic template \texttt{(iso\_P8R3\_SOURCE\_V2\_v01.txt)}\footnote{\url{https://fermi.gsfc.nasa.gov/ssc/data/access/lat/BackgroundModels.html}} for the Galactic and isotropic diffuse emission model, respectively. 
We derive model fluxes of resolved \gray sources from the 4FGL-DR2 catalog\footnote{\url{https://heasarc.gsfc.nasa.gov/W3Browse/fermi/fermilpsc.html}}~\citep{4FGL, 4FGL-DR2}.

First of all, for the UGRB construction, we perform the maximum likelihood analysis in each ROI to fit spectral parameters for all the diffuse and cataloged models above to the photon data.
As described in Ref.~\cite{4FGL-DR2}, the catalog contains point sources with test statistics~(TS) above a detection threshold ($TS \geq 25$) in the 4FGL catalog but below in 4FGL-DR2. In our likelihood run, we exclude flux models of those cataloged sources.
In addition, we perform procedures for the point-source detection in our ROIs.
The photon statistics in our analysis is larger than that in the 4FGL-DR2 catalog, because the observation-time interval in our data is $\sim$20\% longer than one in the 4FGL-DR2 catalog.
Therefore, there should be new point sources of $TS\geq 25$ in our ROIs. 
To detect source candidates, we run the \texttt{find\_sources} method in the \texttt{fermipy} pipeline, applying a flux model of a point-like test source with a power-law spectrum with an amplitude parameter $\alpha$.
To quantify the significance of excess of the candidate flux in the UGRB we constructed in the likelihood run above, we define a TS value as follows ~\citep{10.2307/2957648, 1996ApJ...461..396M},
\begin{align}
& TS \equiv~ 2\Delta \log \mathcal{L}_{\rm max}, \notag \\
& {\rm where}~\Delta \log \mathcal{L} \equiv \log \mathcal{L}(\mathcal{D},\theta|\alpha) - \log \mathcal{L}(\mathcal{D},\theta|\alpha=0).
\label{eq:ts}
\end{align}
$\mathcal{D}$ is the observed photon count. $\theta$ is the best-fit model parameters of the diffuse emission models as well as resolved sources in the 4FGL-DR2 catalog with $TS\geq 25$.
$\Delta \log \mathcal{L}_{\rm max}$ denotes the maximum delta-likelihood when varying the candidate's flux amplitude $\alpha$.
As a result, we find 79 new point sources in ROIs ($\sim 11\%$ of all resolved sources) excluding duplicates in overlapping areas with neighboring ROIs.
After identifying positions of new sources with $TS\geq 25$, we fit their spectral amplitudes and indexes to the UGRB-photon flux.

\begin{figure}
 \begin{center}
  \includegraphics[width=8.5cm]{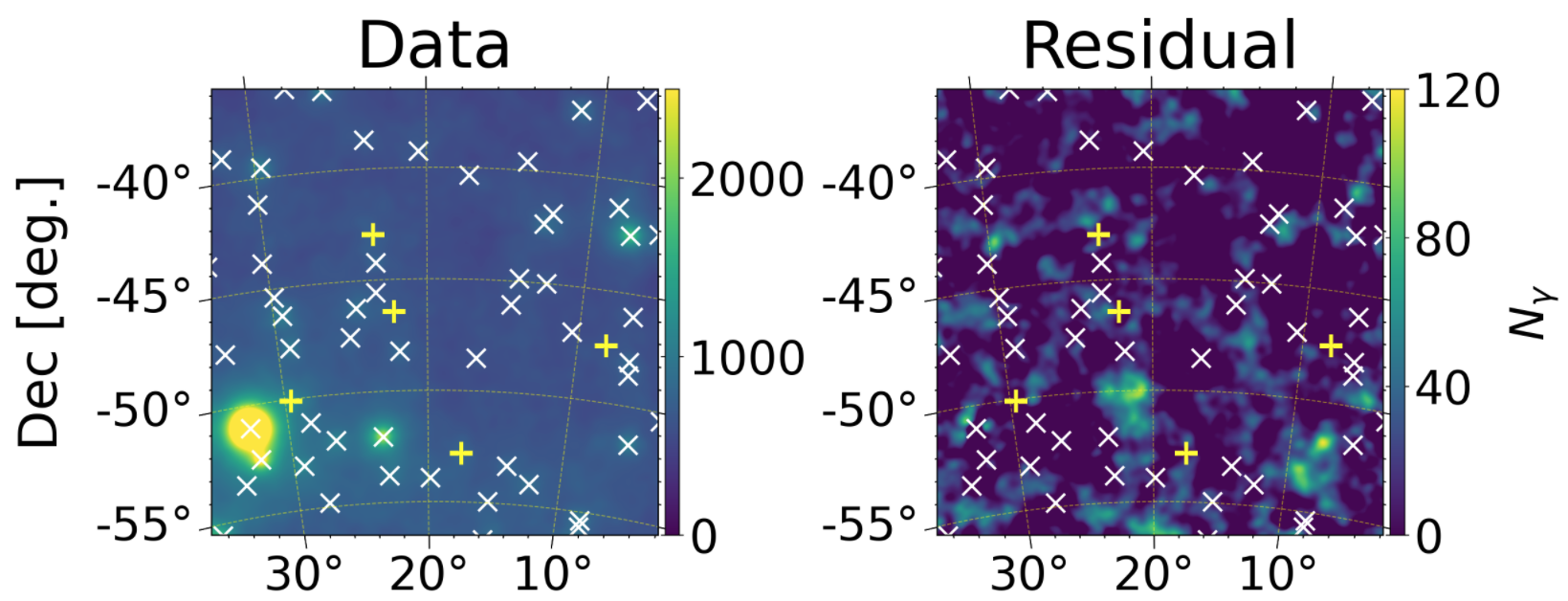}
  \includegraphics[width=8.5cm]{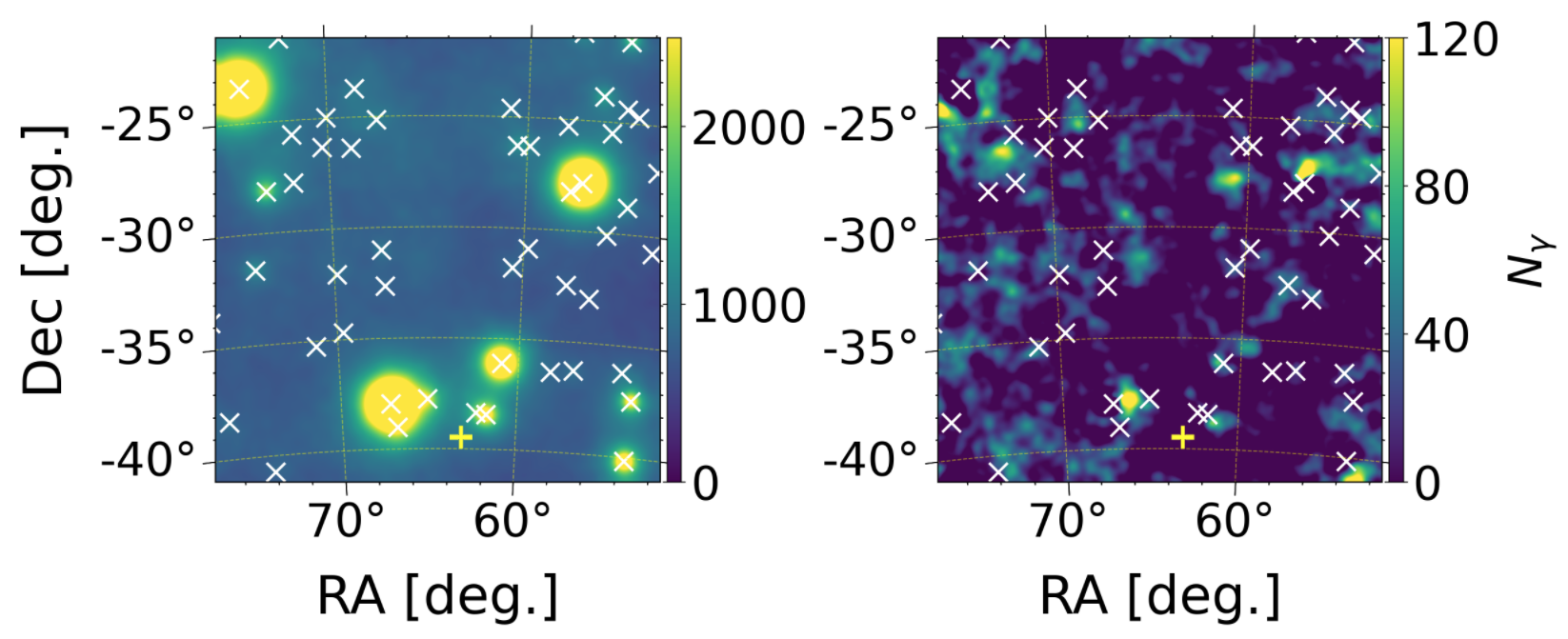}
\end{center}
 \caption{Samples of the photon-count data in two different ROIs~(left panels) and corresponding UGRB maps~(right panels) in the energy range of $500~{\rm MeV} < E_{\gamma} < 500~{\rm GeV}$.
 The cross markers and plus signs represent positions for the 4FGL point sources and for new ones we find, respectively.
 The color scale represents photon counts.
 }
\label{fig:ugrb_sample}
\end{figure}

In Figure~\ref{fig:ugrb_sample}, we show some samples of observed-photon and UGRB maps for different two ROIs, accumulated over all energy bins. The cross and plus markers represent point sources in the 4FGL-DR2 catalog with $TS\geq 25$ and new ones we found, respectively.

\subsection{Flux likelihood profile}
\label{ssec:gammabg}

In this section, we describe a method to estimate the putative \gray flux at the LSBG position in the UGRB field.
For the flux determination of faint sources with TS values of less than 1, the Fermi collaboration has recommended the Bayesian method~\citep{2FGL}.
Accordingly, we estimate the UGRB flux likelihood profile at the LSBG position assuming a power-law flux model with an amplitude parameter $\alpha$, using the delta-likelihood definition in Equation~\ref{eq:ts}.

As mentioned in a relevant work for the UGRB-flux estimation at a faint source position~\citep{2015ApJ...812..159A}, the putative fluxes of LSBGs can be largely affected by the variability of model parameters of bright sources in ROIs as well as neighboring resolved sources.
Therefore, we revisit the UGRB construction around the LSBG position to perform a recalibration for the model parameters in the following steps.
First, for sources with $TS>1000$ in our ROIs, we free all the model parameters.
Also, for all resolved sources within $3^{\circ}$ from the LSBG position, we free all the model parameters for $400<TS\leq1000$ and only the amplitude parameter for $25 \leq TS \leq 400$.
Then, all the freeing parameters are fit to the observed photon data by the \texttt{fit} method in \texttt{Fermipy}.
Note that model parameters except for freed ones are fixed to the values in the UGRB in Section~\ref{ssec:ugrb}.
After all model parameters fixed, we add the LSBG flux model as a simple power-law model with an amplitude parameter $\alpha$ in the UGRB field and obtain the flux likelihood profile by varying $\alpha$.
We find that TS values of the LSBG flux models are lower than 1 in most energy bins.

\section{Methodology}
\label{sec:met}
\ajnvI{In this section, we revisit methodology for the cross-section constraint with a large number of objects without individual redshift. The method follows our previous work~\DvI{\cite{2021arXiv210908832H}} but we introduce the cross covariance between different redshift bins in the $dN/dz$ measurement for more realistic sampling of the redshift distribution.}

\subsection{$dN/dz$ measurement of DES-LSBGs}
\label{ssec:dndz}
\ajnvI{
We measure the redshift distribution of the DES-LSBG sample using the cross-correlation method with the spec-$z$ sample in the same sky.
To extract the clustering information, we measure the angular cross-correlation function between the LSBG sample and the 6dFGS spec-$z$ sample as the spec-$z$ sample. Assuming the linear galaxy bias, we can measure the redshift distribution, $dN_L/dz$.
%
%
%
For measurement of the cross correlation, we use the Landy-Szalay estimator~\citep{Landy-Szalay:1993} with 20 logarithmic angular bins in the range of $0.1 < \theta < 5.0$ [deg], using a open software package \texttt{treecorr (v4.2.3) }~\citep{2015ascl.soft08007J}.
%
%
%
%
Note that we ignore the redshift evolution of the biases within each redshift bin because the redshift range $0<z<0.15$ is narrow. We also 
assume that the evolution of DM clustering within the range is simply scaled with linear-growth factor.}

\begin{figure}
 \begin{center}
  \includegraphics[width=8.5cm]{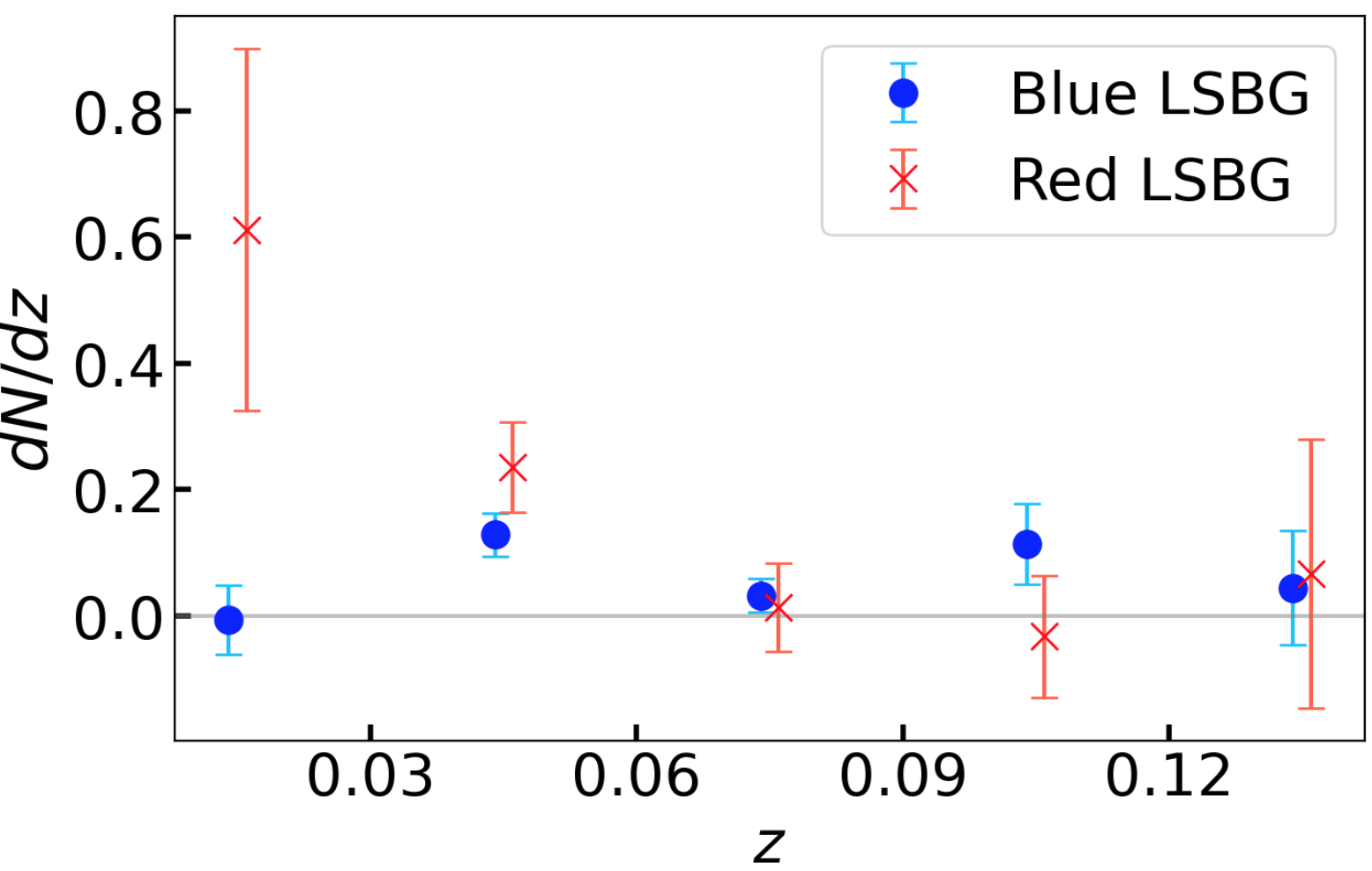}
 \end{center}
 \caption{The measured $dN/dz$ amplitudes for red (cross markers) and blue (circles) LSBGs with 1-$\sigma$ errorbars, \ajnvI{obtained by cross-correlation with the 6dF galaxy sample.}
 }
\label{fig:lsbg_dndz}
\end{figure}

\ajnvI{The statistical uncertainty on the measured $dN/dz$ is evaluated with the 100 jackknife subsamples ($\sim 50$ deg$^2$ for each).}
%
\ajnvI{The full covariance matrix of the projected correlation function can be expressed as
\begin{equation}
    C^{\alpha\beta}_{ij} = \frac{N_{\rm JK}-1}{N_{\rm JK}} \sum_{k=1}^{N_{\rm JK}} [w_{ik}^\alpha - \widehat{w}_{i}^\alpha] [w_{jk}^\beta - \widehat{w}_{j}^\beta],
\end{equation}
where $w^\alpha_{ik}$ is the measured correlation function in the $\alpha$-th redshift and $i$-th angular bin measured on the $k$-th jackknife subsample.
$\widehat{w}^\alpha_i$ is the average of $w^\alpha_{ik}$ over all jackknife subsamples ($N_{\rm JK}$=100).}
%
\ajnvI{
The $dN/dz$ is measured by integrating the correlation function across the angular scales with appropriate weight, here $w=\theta^{-1}$ \citep{2013arXiv1303.4722M}, the covariance can be therefore,}
\begin{equation}
    \bar{C}^{\alpha \beta} = \sum_{ij} C^{\alpha \beta}_{ij} \theta_i^{-1} \theta_j^{-1}.
\end{equation}
We consider that the number of galaxies in each redshift bin is parametrized as free parameters, $\boldsymbol{\Theta}$, obeying multivariate Gaussian distribution with the covariance given in the above equation.
%
%
\ajnvI{To prohibit negative amplitudes of $dN/dz$, we require the step-wise prior, $P(\Theta)=0$ for $\Theta<0$ and $P(\Theta)=1$ otherwise.
We finally apply the linear interpolation of the posterior distribution between each redshift bin and draw random values for every galaxy's redshift from the interpolated posterior distribution.
To be conservative, we set the minimum distance to the objects being 25 Mpc, which is the nearest LSBG found in the HSC-LSBG sample in our previous analysis~\DvI{\citep{2020JCAP...01..059H}}.
In Figure~\ref{fig:lsbg_dndz}, we show the $dN/dz$ for the red and blue LSBG sample with 1-$\sigma$ errorbars evaluated by the error propagation from the jackknife error in the measured angular correlation. 
}

\DvI{We note that for the $dN/dz$ measurement, we have used only LSBG and spec-$z$ samples in the DES Y1 footprint ($\sim 1800~\rm deg^2$) because the random sample of the spec-$z$ sample publicly, used in the angular cross-correlation measurement, is limited to samples within the footprint.}

\subsection{\gray flux modeling for DM annihilation}
\label{ssec:fluxmodel}
We model the \gray flux for the DM annihilation in target objects,
\begin{equation}
\frac{d\Phi_{\rm ann}}{dE} = J \times \frac{\langle \sigma v \rangle}{8\pi m_{\chi}^{2}}
\sum_{i}{\rm Br}_{i}\frac{dN_{i}}{dE},
\label{eq:ann_flux}
\end{equation}
where $\sv$, $m_\chi$, $N_i$ and ${\rm Br}_i$ are the DM annihilation cross-section, DM mass, \gray photon flux and branching ratio of the $i$-th annihilation channel, respectively. In this work, we consider a representative channel, $b\bar{b}$.
$J$ is called as J-factor and defined,
\begin{equation}
J = [1 + b_{\rm sh}(M_{\rm halo})] \int_{s} ds' \int_{\Omega} d\Omega' \rho^{2}_{\rm DM}(s',\Omega'),
\label{eq:j_factor}
\end{equation}
where $\rho_{\rm DM}, s'$ and $\Omega$ denote the DM density, line-of-sight vector and angular size of target objects, respectively.
We assume the NFW profile as the smoothed DM density profile.
\ajnvI{In addition to the smooth component, the annihilation signal is enhanced by the clumpy subhalo structure which is expressed with the boost factor, $b_{\rm st}$. We set $b_{\rm st}$ to unity given the typical halo mass of the LSBG $\sim 10^{10}~\rm M_{\odot}$~\citep{Hiroshima+2018}.}
Because of angular size of DES-LSBGs much smaller than the LAT PSF, we regard the objects as point-like sources, and then J-factor is simply written by the halo mass, the concentration parameter and angular diameter distance.
To estimate the halo mass, we first convert the $g, i$ and $r$ band magnitude into the $V$ band magnitude using~ $V=g-0.59(g-r)-0.01$~\citep{2005AJ....130..873J}. \DvI{Given the luminosity distance $d_L$ by random assignment of object redshifts from the posterior distribution, we obtain the absolute magnitude $M_V$ from $M_V = V+5-5\log_{10}d_L$, and then assuming the mass-to-light ratio to unity~\citep{2008MNRAS.390.1453W}, we finally derive the halo mass from the stellar to halo mass relation~\cite{2013MNRAS.428.3121M}.}
For the concentration parameter, we apply the model~\cite{2019ApJ...871..168D}. 
For statistical uncertainties of the halo mass and density profile at 1-$\sigma$ Gaussian error, we evaluate $\Delta \log M_{\rm halo}=0.4$ for scatter of the stellar-to-halo mass conversion~\citep{2020JCAP...01..059H} and $\Delta \log c=0.1$~\citep{Dutton+:2014}.

\subsection{Composite likelihood}
\label{ssec:composite}
For the flux likelihood of our LSBG, we adopt the Bayesian method as described in section~\ref{ssec:gammabg} 
and 
\DvI{define the likelihood of the annihilation \gray flux model for each LSBG as 
${\mathcal L}(\alpha_{ij}|\sv, J_i)$, where $J_i$ and $\alpha_{ij}$ are the J-factor of $i$-th LSBG and the UGRB amplitude parameter at the object position in the $j$-th energy bin in Section~\ref{ssec:gammabg}, respectively.}
\ajnvI{Given the density of LSBG sample in the sky, }we assume that the flux amplitudes of all LSBGs are \ajnvI{statistically} independent of each other \DvI{\cite{2021arXiv210908832H}}. 
Then, the composite likelihood is given by $\log {\mathcal L}(\alpha|\sv,J) = \sum_{ij} {\mathcal L}(\alpha_{ij}|\sv, J_i)$.
Here, we define a delta-likelihood, $\Delta \log {\mathcal L}(\alpha|\sv,J) \equiv \log {\mathcal L}(\alpha|\sv,J) - \log {\mathcal L}_0$, where ${\mathcal L}_0$ is the likelihood for zero flux of the annihilation \gray flux.
Because of our LSBG flux models to be positive definite, $\Delta \log {\mathcal L}(\alpha|\sv,J)$ are well-described the $\chi^2/2$ distribution with one degree of freedom via the Wilks theorem~\citep{10.2307/2957648}.
Therefore, the upper limit with 95\% C.L. on the annihilation cross-section is given when $\Delta \log {\mathcal L}_(\alpha|\sv,J)$ decreases to 3.8/2.
To evaluate the total uncertainty for the upper-limit estimate, which comes from the $dN/dz$ measurement error and uncertainties for the halo-property estimate, we iterate the Monte-Carlo resampling in 1000 times in terms of the modeling process of the annihilation \gray flux.
For the $dN/dz$ measurement error, we iterate random draw processes for individual object redshifts by using different $dN/dz$ distributions whose amplitudes follow the posterior distribution described in \ref{ssec:dndz}.
For uncertainties of the halo mass and concentration, we give scatters of those Gaussian errors described in section~\ref{ssec:fluxmodel}.

\section{Result}
\label{sec:result}

\begin{figure}
 \begin{center}
  \includegraphics[width=8.5cm]{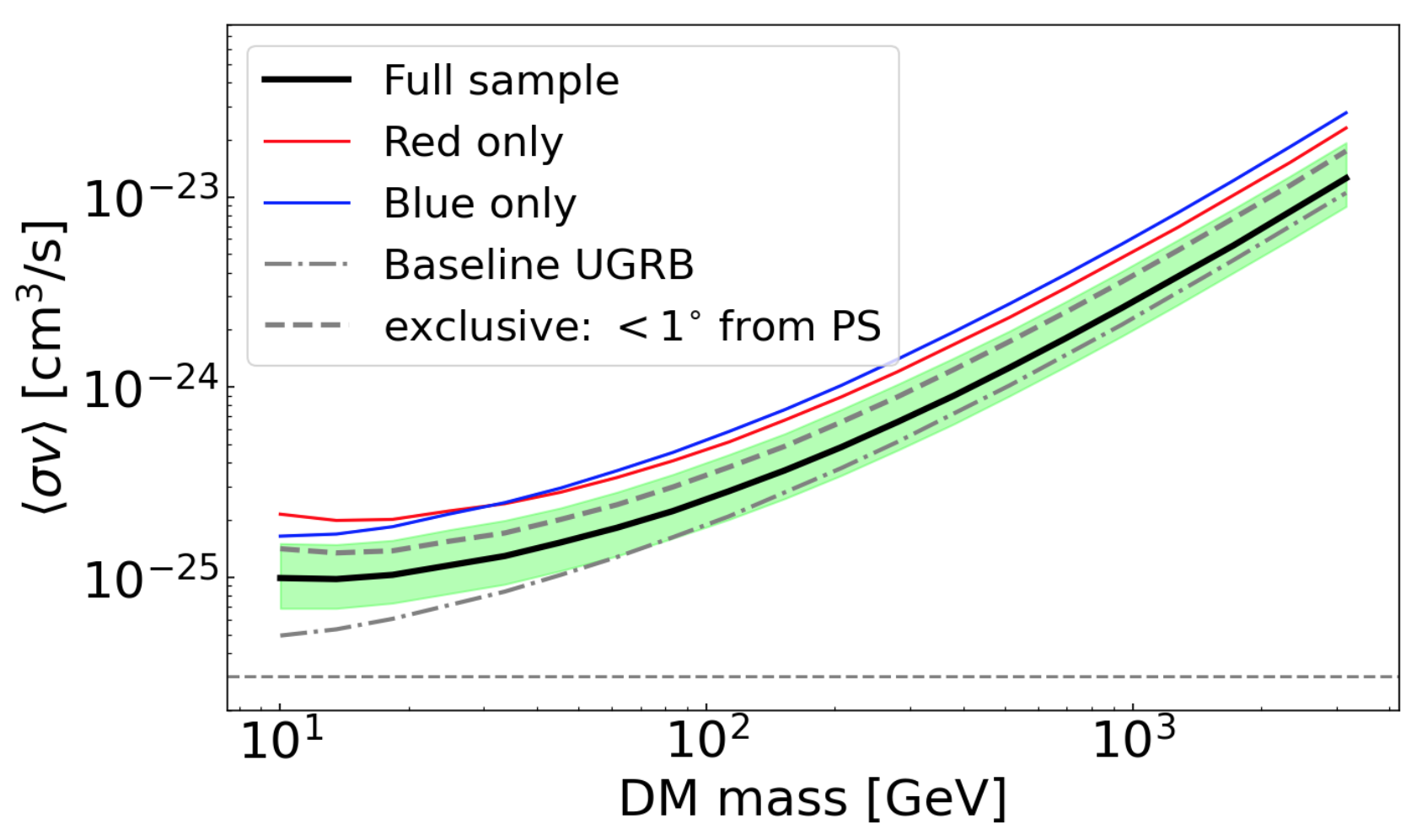}
 \end{center}
 \caption{The upper limits on the velocity-averaged DM annihilation cross-section at the $b\bar{b}$ channel as a function of DM masses.
 \ajnvI{The black, red and blue solid lines are median values of the upper limits on the cross section from the full, red and blue samples of DES LSBGs, respectively. Green shaded region represents a 95\% uncertainty region around the full sample constraint due to the random draw of the redshift.
 Thick dashed line show the constraint from the conservative sample (see text for detail).}
 In the dash-dotted line, we show the constraint with all objects and the UGRB \DvI{without the recalibration process
 described in Subsection \ref{ssec:gammabg}}.
 \ajnvI{The horizontal line represents the prediction of the thermal relic cross section \DvI{($3\times 10^{-26}~\rm [cm^3/s]$)}.}
 }
\label{fig:sigmav_limit}
\end{figure}

In Figure~\ref{fig:sigmav_limit}, we provide upper limits on the annihilation cross-section at the $b\bar{b}$ channel in the black, red and blue solid lines, by the composite likelihood analysis of the annihilation \gray flux models for full, red and blue DES-LSBG samples, respectively.
We show the total uncertainty for the upper limit with the full sample in the green shaded region of 95\% containment in 1000 Monte-Carlo simulations.
Although the number of the blue LSBGs are two times larger than that of the red LSBGs, the upper limits with both populations are similar to each other, reflecting the \ajnvI{fact that the blue LSBGs are distributed widely across all redshift ranges but red LSBGs are localized at lower redshifts.}

\ajnvI{For conservative constraints, we consider the sample removing neighboring objects with separation less than $1^{\circ}$ from bright point sources because UGRB flux around the bright sources may be under estimated due to over-subtracting the model fluxes of the sources, which leads to provide a spuriously strong constraint.}
\ajnvI{In Figure~\ref{fig:sigmav_limit}, we show the constraint from this conservative sample of 16,353 objects as dashed line.
We find that the number of samples due to this selection decreases to $\sim 70\%$ and the constraint becomes $\sim 30\%$ weaker compared to the full sample.
This implies that the constraint simply scales with the number of sample, which suggests that the use of the objects near the bright point sources do not induce spurious constraints in our analysis.
}

The dash-dotted line in Figure~\ref{fig:sigmav_limit} represents the upper limit using the full LSBG sample and UGRB field with the recalibration process in Section~\ref{ssec:gammabg}.
In all DM mass ranges, upper limits with the UGRB from recalibrated model parameters are weaker than those with the UGRB with no recalibration.
As the photon statistics get large in lower energy, inaccurate parameter estimation affects to the likelihood analysis in lower energy and as a result, the upper limits in lower DM mass regimes have relative large variability.

\section{Summary}
\label{sec:summary}

%
\ajnvI{In this work, we place the constraint on the DM annihilation cross-section by cross-correlating the DES-LSBG (23,790 objects in the sky coverage of $\sim 5,000~\rm deg^2$) without individual distance information to the UGRB photon field obtained by the \textit{Fermi}-LAT observation in the energy range of 500~MeV to 500~GeV, accumulated over 12 years.
As performed in our previous work~\cite{2021arXiv210908832H},
the distance of individual galaxy is randomly drawn from the estimated $dN/dz$ distribution, which is measured by the cross-correlation of LSBGs with galaxies whose distances are spectroscopically identified.}
For the $dN/dz$ measurement, we have estimated the angular cross-correlation of all red or blue LSBG samples with different redshift samples of the 6dFGS spec-$z$ sample which is divided into five redshift bins with equal width in the range of $0<z<0.15$.
%
%
\DvI{We have computed the annihilation \gray flux for each LSBG by randomly drawing the redshift from the measured $dN/dz$ distribution.}
%
%

\ajnvI{To obtain the putative flux of each LSBG, we have constructed the UGRB field in our ROIs with the LAT data and all \gray emission models for the diffuse and resolved sources} \DvI{taken from the 4FGL-DR2 source catalog and 79 newly detected sources with $TS\geq 25$ in our analysis, which is $\sim 11\%$ of the number of all resolved sources in our ROIs.
After optimizing all model parameters of these sources in each ROI, we have performed the recalibration for model parameters of the diffuse and resolved sources around LSBGs, because the putative \gray flux, 
\ajnvI{which is typically faint can be highly sensitive to the model parameters of bright sources}
\citep{2015ApJ...812..159A}. 
%
\ajnvI{The putative flux for the LSBG assuming the power-law flux model found to be significantly small because the TS values for most of the LSBGs are less than unity in most of the energy bins.}
%
}

\DvI{
\ajnvI{We have obtained the upper limit of the dark matter cross section of $\sv < 3\times 10^{-25} (95\%{\rm C.L.})$} at $m_\chi=100$ GeV from the composite likelihood for all DES-LSBGs using the recalibrated UGRB flux.
To be conservative, we have removed LSBGs within $1^{\circ}$ radius from resolved sources ($\sim 30\%$ of the total sample) because UGRB fluxes can be underestimated by 
\ajnvI{the contamination from the bright sources,}
which may lead to a spuriously strong constraint.
\ajnvI{However, the amount of the degradation of constraint with this conservative sample selection is consistent with the expectation from the scaling of the number of objects, which indicates that the use of the LSBGs near the bright sources does not induce an unreasonably strong constraint.}
Furthermore, we have considered the impact of the recalibration of the bright-source model parameters on the DM constraint.
We found that the upper limit using the UGRB flux with the recalibration is at most 2 times larger than the one using nonrecalibrated flux. 
%
\ajnvI{Therefore, we conclude that the recalibration of the UGRB fluxes around bright sources is essentially important for this sort of analysis where the estimation of the faint \gray flux is important.}
}


\begin{acknowledgments}
We thank Oscar Macias for giving lots of kind supports,  particularly for the use of the Fermi Science Tool, and productive discussion.
This work was supported in part by World Premier International Research Center Initiative, MEXT, Japan, and JSPS KAKENHI Grant No. 19H00677, 20H05850, 20H05855, 20J11682 and 21H05454.
\end{acknowledgments}


\bibliographystyle{unsrt}
\bibliography{bibdata}

\end{document}